\documentclass[conference]{IEEEtran}
\ifCLASSINFOpdf
\else
\fi
\usepackage{booktabs}
\usepackage{multirow}
\usepackage{balance}
\usepackage[usenames,dvipsnames]{xcolor}
\usepackage[ruled, vlined, linesnumbered]{algorithm2e}
\SetKwRepeat{Do}{do}{while}%
\usepackage{hyperref}
\usepackage{graphicx}
\usepackage{caption}
\usepackage{subcaption}
\begin{document}
\newcommand{\todoc}[2]{{\textcolor{#1}{\textbf{#2}}}}
\renewcommand{\todoc}[2]{\relax}

\newcommand{\todored}[1]{\todoc{red}{\textbf{[[#1]]}}}
\newcommand{\todogreen}[1]{\todoc{ForestGreen}{\textbf{[[#1]]}}}
\newcommand{\todoblue}[1]{\todoc{blue}{\textbf{[[#1]]}}}
\newcommand{\todogray}[1]{\todoc{Gray}{\textbf{[[#1]]}}}
\newcommand{\todoorange}[1]{\todoc{YellowOrange}{\textbf{[[#1]]}}}

\newcommand{\TODO}[1]{\todored{TODO: #1}}
\newcommand{\lin}[1]{\todogreen{Lin: #1}}
\newcommand{\jinqiu}[1]{\todoblue{Jinqiu: #1}}
\newcommand{\annie}[1]{\todoblue{Annie: #1}}
\newcommand{\changed}[1]{{\textcolor{red}{#1}}}
\newcommand{\erik}[1]{\todoblue{Erik: #1}}

\newcommand{\breakc}[2][l]{%
  \begin{tabular}[#1]{@{}l@{}}#2\end{tabular}}

\newcommand{\toolS}{D2Spec\space}
\newcommand{\tool}{D2Spec}
\newcommand{\numAPIsGuru}{236\space}
\newcommand{\numAPITotal}{745\space}
\newcommand{\numAPIHarmony}{681\space}
\newcommand{\numOtherAPI}{44\space}
\newcommand{\numAPIsGuruEvaluated}{64\space}
\newcommand{\numAPIsTotal}{120\space}
\newcommand{\numGoogleAPI}{20\space}

\newcommand{\basePrecisionTotal}{87.5\%}

\newcommand{\pathDocTotal}{2,393}
\newcommand{\pathToolTotal}{2,394}
\newcommand{\pathToolTotalCorrect}{1,930}
\newcommand{\pathPrecisionTotal}{81.3\%}
\newcommand{\pathRecallTotal}{80.6\%}

\newcommand{\methodDocTotal}{2,234}
\newcommand{\methodToolTotal}{2,017}
\newcommand{\methodToolTotalCorrect}{1,703}
\newcommand{\methodPrecisionTotal}{84.4\%}
\newcommand{\methodRecallTotal}{76.2\%}

\newcommand{\compactItem}[1]{\smallskip\noindent\textbf{#1}}
\newcommand{\compactHeader}[1]{\medskip\noindent\textbf{#1}\smallskip\\\noindent}
\newcommand{\compactInlineHeader}[1]{\medskip\noindent\textbf{#1}}


\title{Automatically Extracting Web API Specifications from HTML Documentation}

%
\author{\IEEEauthorblockN{Jinqiu Yang\IEEEauthorrefmark{1},
Erik Wittern\IEEEauthorrefmark{2},
Annie T.T. Ying\IEEEauthorrefmark{3}\IEEEauthorrefmark{2},
Julian Dolby\IEEEauthorrefmark{2},
Lin Tan\IEEEauthorrefmark{1}}
\IEEEauthorblockA{\IEEEauthorrefmark{1} University of Waterloo, Waterloo, Canada}
\IEEEauthorblockA{\IEEEauthorrefmark{2} IBM Research, Yorktown Heights, NY, USA}
\IEEEauthorblockA{\IEEEauthorrefmark{3} EquitySim, Vancouver, Canada}
j223yang@uwaterloo.ca, witternj@us.ibm.com, annie.ying@gmail.com, dolby@us.ibm.com, lintan@uwaterloo.ca
}


\maketitle

\begin{abstract}
Web API specifications are machine-readable descriptions of APIs. These specifications, in combination with related tooling, simplify and support the consumption of APIs.
However, despite the increased distribution of web APIs, specifications are rare and their creation and maintenance heavily relies on manual efforts by third parties.
In this paper, we propose an automatic approach and an associated tool called \toolS for extracting specifications from web API documentation pages.
Given a seed online documentation page on an API, \toolS first crawls all documentation pages on the API, and then uses a set of machine-learning techniques to extract the base URL, path templates, and HTTP methods -- collectively describing the endpoints of an API.

We evaluated whether \toolS can accurately extract endpoints from documentation on $120$ web APIs. The results showed that \toolS achieved a precision of \basePrecisionTotal~in identifying base URLs, a precision of \pathPrecisionTotal~and a recall of \pathRecallTotal~in generating path templates, and a precision of \methodPrecisionTotal~and a recall of \methodRecallTotal~in extracting HTTP methods. 
In addition, we found that \toolS was useful when applied to APIs with pre-existing API specifications:  \toolS revealed many inconsistencies between web API documentation and their corresponding publicly available specifications.  Thus, \toolS can be used by web API providers to keep documentation and specifications in synchronization.
\end{abstract}


%
\IEEEpeerreviewmaketitle
\section{Introduction}
\label{sec:introduction}
Web Application Programming Interfaces (web APIs or simply \emph{APIs} from hereon) provide applications remote, programmatic access to resources such as data or functionalities.
For application developers, the proliferation of such APIs provides tremendous opportunities. Applications can take advantage of vast amount of existing data, like location-based information from the Google Places API\footnote{\url{https://developers.google.com/places}}, hook into established and global social networks, using for example Facebook's\footnote{\url{https://developers.facebook.com/docs/graph-api}} or Twitter's\footnote{\url{https://developer.twitter.com/en/docs/api-reference-index}} APIs, or outsource critical and hard to implement functionalities, like payment processing using the Stripe API.\footnote{\url{https://stripe.com/docs/api}}

To consume APIs, though, developers face numerous challenges~\cite{Wittern:2017b}:
The need to find and select the APIs meeting their requirements, both from a functional and non-functional point-of-view~\cite{Wittern:2016}.
They need to familiarize with the capabilities provided by an API and how to invoke these capabilities, which typically involves studying HTML-based documentation pages that vary across APIs. Compared to using library APIs, for example of a Java library, consumers of web APIs do not have interface signatures readily available or accessible via development tools.
In addition, web APIs are under the control of independent providers who can change the API in a way that can break client code~\cite{Li:2013,Espinha:2015}.
Even for supposedly standardized notions such as the APIs' URL structures, HTTP methods, or HTTP status codes, the semantics and implementation styles differ across APIs~\cite{Rodriguez:2016}.

One attempt to mitigate these problems is to describe APIs in a well-defined way using web API \emph{specifications}.\footnote{
  \annie{To Julian: Can you please review this footnote}We acknowledge that the term \emph{specification} relates to a much more comprehensive description of an application's or system's syntax \emph{and} semantics.
}
  Web API specification formats, like the OpenAPI Specification~\cite{swagger} or the RESTful API Modeling Language (RAML)~\cite{RAML}, describe the URL templates, HTTP methods, headers, parameters, and input and output data required to interact with an API. Being machine-understandable, web API specifications are the basis for various tools that support API consumption:
specification are input for generating consistent API documentation pages\footnote{e.g., \url{https://editor.swagger.io/} or \url{https://github.com/Rebilly/ReDoc}},
they are used to catalog APIs\footnote{e.g., \url{https://apiharmony-open.mybluemix.net/} or \url{https://any-api.com}},
to auto-generate software development kits that wrap APIs in various languages\footnote{e.g., \url{https://swagger.io/swagger-codegen} or \url{https://apimatic.docs.apiary.io}},
or even to statically check client code for possible errors~\cite{Wittern:2017}.

Unfortunately, client developers cannot leverage these advantages from such tooling unless web API specifications are available.
In contrast to over 19 thousand APIs listed in ProgrammableWeb~\cite{ProgrammableWeb} (an authoritative directory of web APIs referred as ``The Journal of the API Economy''\footnote{https://techcrunch.com/2013/08/07/veteran-news-editor-david-berlind-joins-programmable-web-the-journal-of-the-api-economy/}), fewer than a thousand specifications are publicly available on APIs.guru~\cite{apiguru}, the largest publicly available directory of web API specifications.
\annie{To Eric: Do we know how many of the APIs.guru's specs are specs that from the API provided and how many are generated?}

\erik{Annie, unfortunately information on how many specs are made available by API providers is not available. I propose to fall back to stating that APIs.guru depends on 1) manual efforts by the community and 2) API-specific scripts to generate and maintain specs (see also commented-out text above).}

\begin{figure}[!htb]
  \centering
  \includegraphics[width=\columnwidth]{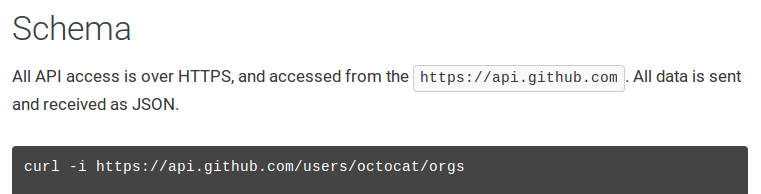}
  \caption{Example-style documentation (GitHub API)}
  \label{fig:github_doc_example}
\end{figure}

\begin{figure}[!htb]
  \centering
  \includegraphics[width=0.8\columnwidth]{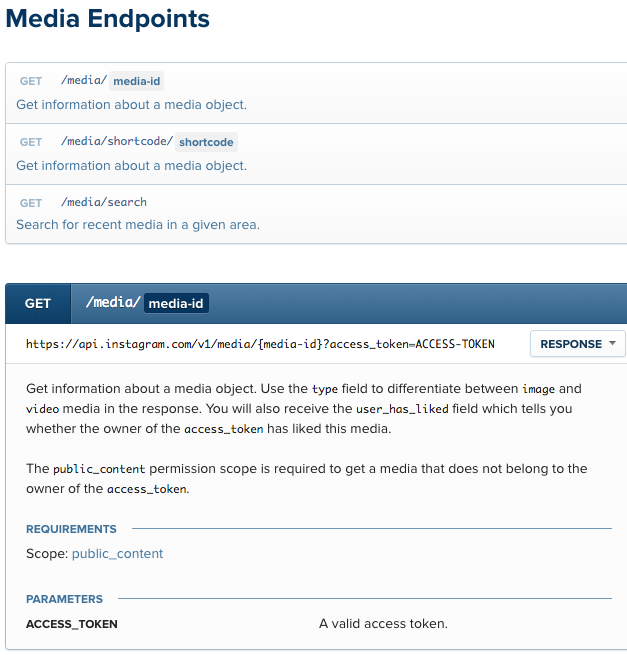}
  \caption{Reference-based API documentation (Instagram API)}
  \label{fig:instagram_doc_example}
\end{figure}

The goal of this research is to provide client developers or API catalogs access to a much larger number of specifications by extracting them from much more prevalent, semi-structured online documentation (typically in form of HTML pages).
Many software engineering researchers have looked into a similar problem but in the traditional library API context, namely, identifying Java method signatures from API documentation~\cite{recover2012,subramanian2014live,rigby2013discovering}.
These approaches share the assumption that method signatures or code elements being extracted are written in Java, adhering to the specific Java grammar and conventions. Approaches capable of extracting web API endpoint descriptions not only require an adjustment to the search pattern for web API endpoints, but also need to account for the two common but distinct styles of web API documentation:
an example-based style (e.g., the GitHub API documentation as shown in Figure~\ref{fig:github_doc_example} uses an example-based style, where the base URL \texttt{https://api.github.com} and the path template \texttt{/users/ \{username\}/orgs} are embedded in free-form text and a \texttt{curl} command) and a more structured, reference-based documentation style (e.g., the Instagram API, Figure~\ref{fig:instagram_doc_example}).

In this paper, with these two distinct styles of web API documentation in mind, we propose an approach, and the associated tool \tool, to automatically extract a web API specification -- more specifically, endpoint descriptions consisting of a base URL, path templates, and HTTP methods -- from API documentation pages containing free-form text and arbitrary HTML structures.
Given a seed documentation page for an API, \toolS first crawls the associated documentation pages and then uses a set of machine learning techniques to extract the base URL of an API (e.g., \texttt{https://api.github.com}), the path templates (possibly containing path parameters, e.g., \texttt{/users/\{user\-name\}/org}), and HTTP methods (e.g., \texttt{GET}, \texttt{POST}).
More specifically, \toolS uses classifiers and a hierarchical clustering algorithm to extract a base URL and path templates for an API, and searches the context of a path template to infer the HTTP method.

We evaluated whether \toolS can accurately extract endpoints from documentation on $120$ web APIs. The results showed that \toolS achieves a precision of \basePrecisionTotal~in identifying base URLs, a precision of \pathPrecisionTotal~and a recall of \pathRecallTotal~in generating path templates, and a precision of \methodPrecisionTotal~and a recall of \methodRecallTotal~in extracting HTTP methods. 
In addition, from an evaluation on \numAPIsGuruEvaluated APIs with pre-existing API specifications, we found that \toolS was also useful: \toolS revealed many inconsistencies between web API documentation and their corresponding publicly available specifications. Thus, \toolS can be used by web API providers to keep documentation and specifications in synchronization.

\toolS currently does not infer the structure of data being sent to or received from an API, nor HTTP headers. However, 
since most requests use the \texttt{GET} method~\cite{Rodriguez:2016} and thus do not expect any request payload, in those cases, the combination of a base URL and path template already allow for a successful API invocation.
Extending our work to extract the structure of data, for example through schema inference of provided example response data, is future work.

In the remainder of this paper, we present our approach to extract API specifications from documentation using a combination of machine learning techniques in Section~\ref{sec:approach}. We present an empirical evaluation in Section~\ref{sec:evaluation}.  
We discuss threats in Section~\ref{sec:threats} and related work in Section~\ref{sec:related_work} before concluding in Section~\ref{sec:conclusion}.



\section{Approach}
\label{sec:approach}
In this section, we describe how \toolS combines machine learning classification and hierarchical clustering to extract significant parts of web API specifications from online documentation. 
\toolS focuses on extracting three components of API specifications: base URLs, path templates, and descriptions (i.e., HTTP methods).

The web API's \textbf{base URL} is essential in a web API specification: any URL of a Web API request must contain the base URL and the relative path of the corresponding endpoint.
More formally, a base URL is a common prefix of all URLs for web API invocations, excluding other URLs such as documentation pages.
In OpenAPI specifications, a base URL is constructed via three fields: a \emph{scheme} (e.g., \texttt{https}), the \emph{host} (e.g., \texttt{api.in\-stagram.com}), and optionally a \emph{base path} (e.g., \texttt{/v1}).
In many APIs (e.g., \emph{Instagram API}), the base URL is the \emph{longest} common prefix of all the URLs for invoking the web API.  However, for other APIs, such as Microsoft's \emph{The DevTest Labs Client API}, the longest common prefix is \texttt{https://management. azure.com/subscriptions} while the actual base URL is \texttt{https://management.azure.com}, because \texttt{/subscriptions} is defined to be part of the endpoint paths.  Whether a base URL is indeed the longest common prefix is a design decision of the API provider.

A \textbf{path template} defines fixed components of a URL as well as ones to be instantiated dynamically. For example, in the path template \texttt{/users/\{userId\}/posts}, the part \texttt{\{userId\}} is a \emph{path parameter} that needs to be instantiated with a concrete value of a user ID before performing a request. 
A path parameter is typically denoted via enclosing brackets (i.e., ``\{\}'', ``[]'', ``$<>$'', or ``()'') or a prefix ``:''.

\toolS focuses on one type of \textbf{description} associated with the path template: the \textbf{HTTP method}. It reflects the type of interaction to be performed on a resource exposed by a web API. While many web APIs long relied on \texttt{GET} and \texttt{POST}, now a much broader spectrum of methods is used~\cite{Rodriguez:2016}. As proposed in related work, we denote every valid combination of a path template and an HTTP method as an \emph{endpoint} of the API~\cite{suter2015inferring}.

\begin{figure*}
  \centering
  \includegraphics[scale=0.47]{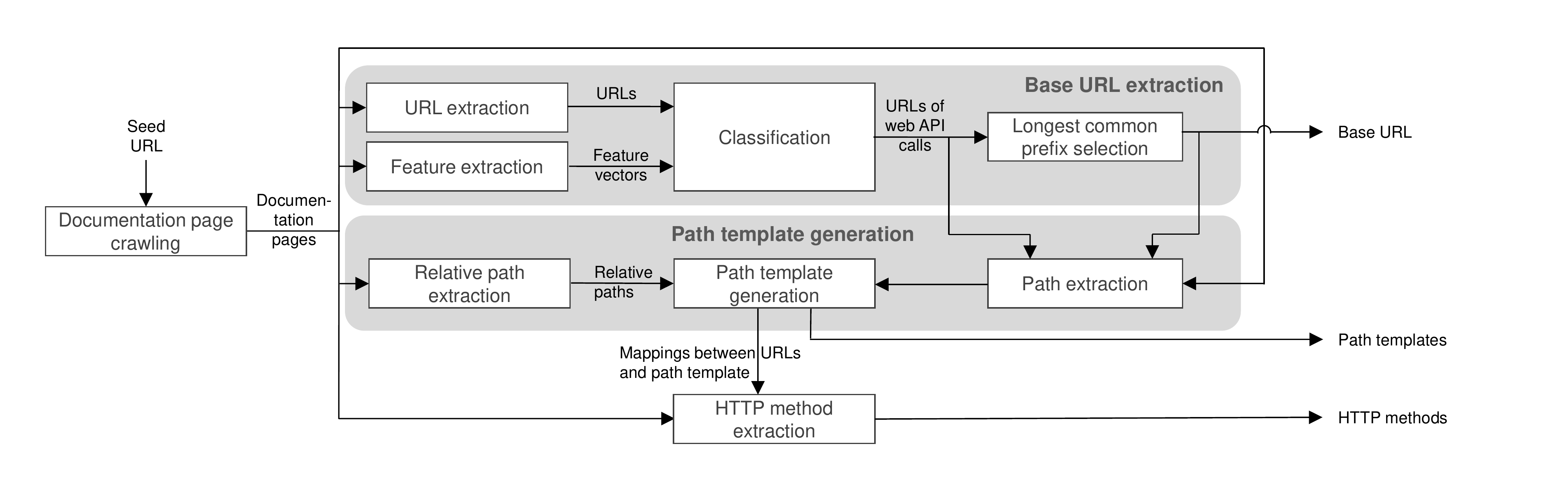}
  \vspace{-0.1in}
  \caption{Overview of \tool, consisting of `Base URL Generation' and `Endpoint Generation'. \toolS takes web API documentation as input and outputs specifications that include base URLs and endpoints.\label{fig:overview}}
\end{figure*}

\toolS combines a set of techniques to infer the base URL, path templates, and HTTP methods, given a seed documentation page.
Figure~\ref{fig:overview} provides an overview.
In the first step, \toolS uses a simple crawling approach to obtain complete documentation sources for an API. The crawler, starting from the provided seed page, iteratively downloads all linked \emph{sub pages}. For crawling, \toolS uses the headless browser Splash\footnote{\url{https://scrapinghub.com/splash/}} to execute any JavaScript on each page before downloading it, as this may impact the resulting HTML structure.
In order to extract the \textbf{base URL} (see Section~\ref{sec:approach_baseurl}), \toolS first extracts all candidate URLs that can represent a web API call from the crawled documentation pages;
\toolS next leverages machine learning classification to determine for each candidate URL whether or not it is likely to represent an invocation to the documented web API.  Finally, \toolS selects the longest common prefix of these URLs.
For the \textbf{path templates} (see Section~\ref{sec:approach_paths}), \toolS leverages the URLs likely to be invocations of the API and extracts the URL paths (the part of the URL after removing the base URL). From these paths, using an agglomerative hierarchical clustering algorithm, \toolS infers path templates by identifying path parameters and aggregating paths.
\toolS then finds the descriptions co-located with the URL paths in the documentation, from which it extracts the \textbf{HTTP method(s)} (see Section~\ref{sec:approach_methods}) that can be combined with each path template (forming endpoints). 


\subsection{Base URL Extraction}
\label{sec:approach_baseurl}
Identifying the base URL in online documentation is not as straightforward as searching for keywords or templates such as ``The base URL is $<$\emph{base URL}$>$''; 
base URLs are often not explicitly mentioned in the documentation.
Rather, base URLs are often included as part of depicted examples of web API requests, as in the case of the GitHub documentation shown in Figure~\ref{fig:github_doc_example}.
Thus, \tool's approach is to infer the base URL from all the URLs provided in online documentation.

\compactHeader{Step 1 - Extracting URLs}
As a first step, \toolS extracts all candidate URLs in the documentation that represent web API calls.
These candidate URLs consist of standard URLs (according to the W3C definition~\cite{url_definition}) and URLs containing path parameters enclosed in ``\texttt{\{\}}'', ``\texttt{[]}'', ``\texttt{()}'', or ``\texttt{<>}''.
We do not include in this list URL links within \texttt{href} attributes of link tags, nor inside \texttt{<script>} tags: URLs that represent web API calls are one of the main content in a documentation page to be communicated to the readers; hence, such URLs tend to be rendered in the documentation rather than appear as links or in scripts.
Even excluding such links, some of the URLs in the candidate list may not represent web API calls, e.g., URLs of related or even unrelated resources. In fact, we studied a set of $15$ web APIs\footnote{This data set of $15$ web APIs, which we studied  to design \tool, is independent of our evaluation set.} and found that 42\% of the contained URLs are \emph{not} invocations of web APIs.

\compactHeader{Step 2 - Identifying URLs of web API calls}
To filter out spurious URLs that do not represent web API calls, we use supervised machine learning classification to determine whether each URL from Step 1 is likely to represent a web API call.
For each URL,
\toolS generates two categories of features: properties of documentation pages and properties of the URL itself. 
The first category consists of four features related to properties of the documentation pages from which URLs were extracted:
\begin{itemize}
  \item{\texttt{clickable}}: True, if the URL is part of the link text enclosed in \texttt{<a>} tags with the ``\texttt{href}'' attribute.

  \item{\texttt{code\_tag}}: True, if the URL appears inside {\tt <code>} tags, and false otherwise.

  \item{\texttt{within\_json}}: True, if the URL is inside valid JSON within a pair of matched HTML tags.

  \item{\texttt{same\_domain\_with\_doc\_link}}: True, if the URL has the same host name as the URL of the documentation page itself, and false otherwise.
\end{itemize}

\noindent
The second category consists of four features related to properties of the URL itself:
\begin{itemize}
  \item{\texttt{query\_parameter}}: True, if the URL contains query parameters which are denoted by \texttt{?} and/or \texttt{=}. For example, in the URL {\tt https://api.github.com/...\allowbreak/issues?state=closed}, \texttt{state} is a parameter with the value \texttt{closed}. URLs with parameters are more likely to depict web API calls.

  \item{\texttt{api\_convention}}: The number of conventions exhibited by the URL indicates whether it likely corresponds to a web API call. We include three conventions described in previous work~\cite{Rodriguez:2016}:
  (1) whether the URL contains the term \texttt{rest};
  (2) whether the URL contains the term \texttt{api}; and
  (3) whether the URL contains version related information, including the terms \texttt{v[0-9\textbackslash.]+} or \texttt{version[0-9\textbackslash.]+}. For example, if a URL satisfies all three conventions, the value of this feature is 3.

  \item{\texttt{path\_template}}: True, if the URL contains a path parameter denoted by enclosing brackets (\texttt{\{\}, [], (), <>}) or a colon prefix (\texttt{:}).

  \item{\texttt{curl\_return}}: A categorical feature representing the return value from invoking a \texttt{curl} command on the URL.  We determine into which of the following categories the return value of the command falls:
(1) is in JSON format (the URL likely corresponds to a web API request);
(2) contains authentication errors\footnote{We defined an authentication error to be indicated by a HTTP status code of either 401 or 407, or by a return message that contains the string ``Invalid certificate''.} (the URL may correspond to a web request without the proper authentication);
(3) everything else (e.g., in XML/HTML format which likely corresponds to learning resources as opposed to web API requests).

\end{itemize}

We built the classification model based on $15$ web API documentation. This data set is independent of our evaluation data set. We manually identify a base URL for each web API in order to label whether each URL is indeed one web API call in the training set. Note that this manual process is only needed in order to build the training set; we do not need to manually identify base URLs when we apply \toolS on web API documentation. In our evaluation, URLs are automatically labeled by machine learning classification. The feature vector of the URLs is automatically created by \tool.

\toolS uses the support vector machine (SVM) classifier from scikit-learn~\cite{scikit} with the default parameters, both to train the model and to use the trained model in the evaluation as well. The trained model achieves an accuracy\footnote{Accuracy is the percentage of correctly labeled (whether are indeed web API calls or not) URLs out of all the URLs extracted from the documentation.} of $0.97$ and a F1-measure of $0.97$ from the 10-fold cross-validation.

\compactHeader{Step 3 - Extracting longest common prefix}
Finally, \toolS identifies the base URL by computing the longest common prefix of the URLs classified as likely depicting calls to the web API from step 2. This approach is based on the assumption that web API requests are the most frequent type of URLs rendered in the documentation. This assumption works well in practice, as the reported results in Section~\ref{sec:evaluation} show. This step is necessary because although the classification achieves high accuracy (97\%), there are still URLs that do not target the web API.

\subsection{Path Template Generation}
\label{sec:approach_paths}
Having identified an API's base URL as described in Section~\ref{sec:approach_baseurl}, we can use it to extract the path templates of the API.

Paths of endpoints are typically presented in a documentation page in one of two ways:
Absolute URLs describe the whole URL used to perform an API request, for example, \texttt{\footnotesize https://api.github.com/repos/vmg/redcarpet/issues}.
\\When identifying base URLs, \toolS already extracts absolute URLs and can obtain paths of endpoints by truncating the already determined base URL.
Alternatively, documentation pages may only provide relative path components without the base URL, for example \texttt{/users/repo}.
In this case, \toolS extracts relative paths based on heuristics (i.e., using regular expressions). In the experiment, we did not observe a significant number of false endpoint paths caused by this approach. From a manual analysis, we found that unlike URLs, which often include links to external resources, relative paths often describe API endpoints, since they are otherwise not very meaningful to a human. 

In addition, path parameters are denoted in two ways:
A path parameter can be denoted \emph{explicitly} via enclosing syntactic constructs (e.g., ``\texttt{\{\}}'', ``\texttt{[]}'', ``\texttt{<>}'', or ``\texttt{()}'') or by prefixing a path parameter using ``\texttt{:}''.
Other documentation pages \emph{implicitly} indicate path parameters via an example style (e.g., the \emph{GitHub} example in Figure 1), with URLs where parameters are instantiated. For example, in the URL \texttt{\footnotesize https://api.github.com/users/alice/gists}, ``alice'' is an instantiated value of the path parameter \texttt{\{userId\}}. 
Identifying path parameters expressed syntactically is straight-forward, while identifying path parameters in the example based documentation pages requires an algorithm that can determine from the path examples which of the path segments are instantiated values.

\begin{algorithm}[t]
\SetKwProg{Fn}{Function}{}{}
\SetAlgoLined
\KwIn{$paths$ /*a set of paths that represent endpoints*/}
\KwIn{$T$ /* Threshold for merging clusters */}
\KwOut{$c_1, ..., c_n$ /*each cluster $c_i$ groups the paths invoking the same endpoint*/}
\Fn{hierarchical\_clustering (T, paths)}{
$C \gets$ make each path a singleton cluster\\
\Do{$|C| > 1 \wedge progress$}{
  $progress \gets false$\\
  \ForEach{$c_i,c_j\in C$ with min $dist(c_i,c_j)$}{%
    \If{$dist(c_i,c_j) < T$} {
      $progress \gets true$\\
      $C \gets C - \left\{c_i, c_j\right\} \cup \left\{merge\left(c_i,c_j\right)\right\}$
    }
  }
}
}
\caption{Clustering algorithm\label{clustering}}
\end{algorithm}

We propose an iterative algorithm to infer whether a path segment is a fixed segment of an endpoint, a path parameter, or an instantiated value. The algorithm consists of two main ideas. First, it uses clustering to group paths that we infer to invoke the same endpoint. For example, if we found four paths in the documentation for an API:\\
\texttt{/users/\{username\}/repos} \\
\texttt{/users/alice/repos} \\ 
\texttt{/users/alice/re\-ceived\_events} \\
\texttt{/users/bob/received\_events} \\
the clustering algorithm groups the first two into one cluster and the last two into the second cluster.
From the first cluster, we know that \texttt{alice} is an instantiated value of \texttt{\{username\}}.
Second, in subsequent iterations, the algorithm then leverages the fact that \texttt{alice} is an already inferred instantiated value to improve the clustering in the next iteration, marking both \texttt{alice} and \texttt{bob} as two instantiated values.

\toolS uses hierarchical agglomerative clustering~\cite{manning2009introduction}, as described in Algorithm~\ref{clustering}.
Given a set of paths with the same number of segments, the goal is to group the paths so that paths in a cluster invoke the same endpoint. 
We begin with one data point (i.e., one path) per cluster (line 2 in
Algorithm~\ref{clustering}). At each iteration (lines 4-10), we calculate the
distance among all the pair-wise clusters and picks the pair with the
shortest distance (line 5) to merge (line 8). For our implementation, the
distance function (Algorithm~\ref{distance}) considers two paths the
``closest'' if they have exactly the same segments -- each matching
concrete (i.e. not a path parameter) segment $i$ gets one point (Algorithm~\ref{distance}, line 8). Because two paths can never invoke the same endpoint when they have a different number of segments, the distance of such a pair is infinite (Algorithm~\ref{distance}, line 5). If the $j$-th segment of a path is a path parameter, the distance function considers the segment a match on the $j$-th segment of any other paths of the same length, with a discounted point of 0.8 instead of 1 (Algorithm~\ref{distance}, line 8). The clustering algorithm stops when the next pair of clusters to merge has the distance larger than a threshold \emph{T} (Algorithm~\ref{clustering}, lines 6, 7, and 11).  In our implementation, the threshold is set to 1, meaning that we allow paths in a cluster to have a single path segment different from each other. 

\begin{algorithm}[t]
\SetKwProg{Fn}{Function}{}{}
\SetAlgoLined
\Fn{dist(cluster $c_1$, cluster $c_2$)}{
return $\min\limits_{path_1 \in c_1, path_2 \in c_2} dist\_singles(path_1, path_2)$\\
}
\hrule 
\Fn{dist\_singles(list of segments $S_1$, list of segments $S_2$)}{
  \If{$|S_1| \neq |S_2|$}{
    return $\infty$\\
  }\Else{
    $sim \gets \left(\begin{array}{l}
                       \left|\left\{ i \left|concrete(S_1[i])
                       \wedge S_1[i]=S_2[i]\right.\right\}\right| + \\
                       0.8 \times \left|\left\{ i \left| param(S_1[j])
                       \vee
                       param(S_2[j])\right.\right\}\right|\end{array}\right)$\\ 

    $d \gets |S_1| - sim$\\
    return $d$
  }
}
\caption{Distance functions\label{distance}}
\end{algorithm}

To leverage the instantiated values that are already inferred, such as
\texttt{alice} in the \texttt{received\_events} cluster in the
example, we adapt the standard hierarchical agglomerative algorithm as follows (Algorithm~\ref{algorithm}): The algorithm keeps track of a list of instantiated values of the path parameters per API (line 9), and stops when no additional instantiated values are found from the function \emph{infer\_parameter\_value} (lines 10 and 12).  Each iteration starts by updating the paths with the currently known instantiated values (lines 5-7).  These paths are the input to the hierarchical agglomerative clustering algorithm (line 8). Clustering is performed after updating the newly instantiated values because when new path parameters are identified, the similarities will be updated.  Within each cluster, new values of path parameters are inferred (line 10, the call to \emph{infer\_parameter\_value}).  This adapted algorithm can correctly cluster the four paths into two endpoints: \texttt{/users/\{username\}/repos} and \texttt{/users/ \{username\}/received\_events}.

\begin{algorithm}[t]
\SetKwProg{Fn}{Function}{}{}
\SetAlgoLined
\KwIn{$paths$ /*a set of paths that represent endpoints*/}
\KwIn{$T$ /* Threshold for merging clusters */}
\KwOut{$paths$ /*a set of paths with locations of path parameters identified*/}
\SetKwRepeat{Do}{do}{while}
$values \gets \emptyset$ /*the set of values of path parameters*/\\
\Do{$prevValueSize \neq \left|values\right|$}{
$prevValueSize \gets \left|values\right|$ \\
\ForEach{path $\in$ paths}{
annotate the segments of $path$ as parameters if they occur in $values$
}
$clusters$ $\gets$ hierarchical\_clustering(paths)\\
\ForEach{cluster $\in$ clusters}{
$values$.addAll(infer\_parameter\_value($cluster$))
}
}
\hrule 
\Fn{infer\_parameter\_value (cluster)}{
$paramValues$ $\gets$ $\emptyset$ \\
\ForEach{ pair ($path$, $path\_param$) $\in$ cluster}{
  $value$ $\gets$ extract the parameter value at the i-th segment in $path$  where the i-th segment in $path\_param$ is a parameter
  $paramValues$.add($value$)
}
  return $paramValues$
}
\caption{Algorithm for inferring path parameters\label{algorithm}}
\end{algorithm}
\subsection{HTTP Methods}
\label{sec:approach_methods}

In web API documentation, the paths (whether or not path parameters are denoted explicitly or not) are typically co-located with other valuable information that \toolS aims to extract, namely valid HTTP methods to use with a path template (\texttt{GET}, \texttt{PUT}, \texttt{DELETE}...). 
We call the context in which this information exists a \emph{description block} of a path template.
In this section, we first describe how we locate the description block associated with a path template, and then how \toolS extracts the HTTP method. 

\begin{figure}
  \includegraphics[width=\columnwidth]{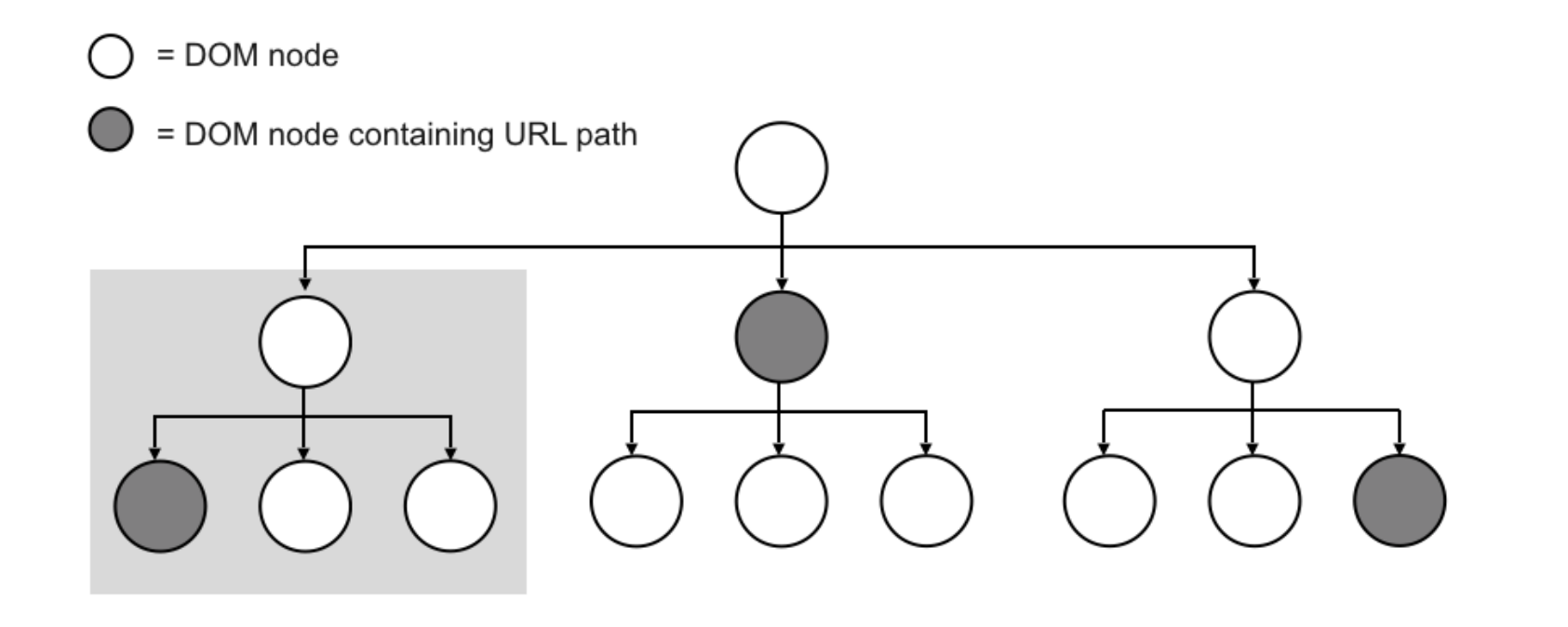}
  \caption{A DOM tree of an HTML page, with each node representing the rendered text from the fragment of the HTML page enclosed in a pair of matched tags. Gray nodes contain a path template. The gray box denotes an example \emph{description block}.
\label{fig:tree}}
\end{figure}

\toolS uses the URLs from the original documentation page that match with the inferred path templates as anchors in documentation pages (in HTML) to locate the scope of the description block.
If there are multiple URLs in the page that match the path template, \toolS combines the contexts of all the URLs as the description block of the path template.  \toolS locates a description block for each path template as follows. 
First, \toolS parses the documentation page into a DOM tree (Figure~\ref{fig:tree}), with each node representing the rendered text from the fragment of the HTML page enclosed in a pair of matched tags.
Second, \toolS marks the nodes whose rendered text contains at least one URL that matches a path template as gray, and locates the description block for each of these nodes. More specifically, for each gray node, \toolS combines its description block by expanding to include (1) the siblings of the node (starting with the closest siblings); then (2) the parent of the node. In (1), assuming the node is the $n$-th child of its parent, the expansion starts from $n-1$ to $0$ and then $n+1$ to the farthest sibling; if a sibling is an ancestor of any other gray node, the expansion terminates entirely. This termination condition applies to 2) as well. The expansion stops if the parent is an ancestor of a node. An example description block is marked by the gray box in Figure~\ref{fig:tree}.

In this work, \toolS focuses on extracting HTTP methods for each path template.

Having determined description blocks, \toolS searches in them for the seven possible method names, namely, \texttt{GET}, \texttt{POST}, \texttt{PUT}, \texttt{DELETE}, \texttt{OPTIONS}, \texttt{HEAD}, and \texttt{PATCH}. If none of these names is found, \toolS uses \texttt{GET} as the default value for method of endpoint, as the GET method is the most popular method for web APIs~\cite{Rodriguez:2016}.

\section{Evaluation}
\label{sec:evaluation}
In this section, we present the evaluation of \tool. We aim to answer two questions: 

\compactInlineHeader{RQ1:} \textit{How accurately can \toolS extract web API specifications from documentation?}
We are interested in a quantitative perspective on how accurately \toolS extracts base URL, path templates, and HTTP methods from the documentation.

\compactInlineHeader{RQ2:} 
\textit{Can \toolS be used to identify inconsistencies between a pre-existing API specification and the API's documentation, pointing to the two being out of synchronization?}
Some web APIs do have the associated specifications readily available. Third parties such as~\cite{apiguru} (Section~\ref{sub:data}) curate specifications for many popular, public APIs. For other public or private APIs, the API providers themselves generate specifications. However, in either case, specifications can become out of synchronization with the documentation if, for example, a provider routinely maintains the one but not the other. We want to evaluate how well \toolS can detect mismatches between documentation and corresponding specifications.

\subsection{Data Collection}
\label{sub:data}
To address the two research questions, we need two types of information.
For RQ1, we need API documentation pages to which we can apply \tool. We can then manually assess how accurate the resulting specifications are as compared to the documentation pages.
To address RQ2, we need APIs that feature both documentation pages and existing specifications. After applying \toolS to the documentation pages, we can assess whether the extracted specifications are in sync with the existing ones. We obtained these two types of information from two sources:

\compactInlineHeader{APIs.guru}~\cite{apiguru} is a third-party effort that curates OpenAPI specifications of popular public APIs. These specifications are either created and maintained through manual community effort, or using API-specific scripts, which translate other specification formats to the OpenAPI specification or are hard-coded to generate specifications from API-specific documentation pages. When we performed the evaluation, APIs.guru hosted \numAPIsGuru specifications, each describing a different API with a unique base URL. Most of these APIs are provided by Google ($127$ APIs) or Microsoft ($22$ APIs), while the remaining $87$ APIs are provided by other individual providers. We found documentation pages for all of the Google APIs, for none of the Microsoft APIs, and for \numOtherAPI of the remaining APIs. If we were to construct the evaluation dataset for the machine learning model by including all APIs for which we found documentation pages, we risk unfairly evaluating the model that biases the Google APIs, whose documentation pages have the same HTML structure while the documentation pages from the other individual providers have different structures among them. Ultimately, we include \numGoogleAPI randomly selected Google APIs in our evaluation. Overall, from APIs.guru, the APIs used in the evaluation consist of \numOtherAPI APIs from the individual providers (we will call this set \emph{GuruIndividual}) and \numGoogleAPI Google APIs (we will call this set \emph{GuruGoogleSample}), thus a total of $64$ APIs.

\compactInlineHeader{API Harmony}~\cite{apiharmony} is a catalog of web APIs that helps developers to find and choose web APIs, and learn how to use them. API Harmony collects information on public web APIs. When we performed the evaluation, API Harmony listed $1,019$ web APIs in total, $772$ of which contained links to the API's documentation page. We crawled the links to these documentation pages with the help of API Harmony's sitemap.xml file.

From these two sources, collectively, we obtained \numAPITotal unique APIs -- $681$ from API Harmony (from $772$, we excluded $91$ APIs
that overlapped with APIs.guru). For RQ1 and RQ2, we will examine a subset of the \numAPITotal as follows:

\begin{itemize}
\item The \emph{GuruIndividual} dataset consists of the \numOtherAPI APIs in APIs.guru that are not from Google or Microsoft.
\item The \emph{GuruGoogleSample} dataset consists of the \numGoogleAPI Google APIs from APIs.guru.
\item The \emph{HarmonySample} dataset consists of a sample of $56$ APIs from the $681$ APIs from API Harmony. Section~\ref{sub:rq1} will describe how this sample was obtained.
\end{itemize}

\subsection{RQ1: Can \toolS accurately extract web API specifications from documentation?}
\label{sub:rq1}

\compactInlineHeader{Approach:} To assess the accuracy of \tool, we aim to determine how well the produced specifications match the input online documentation. 


To increase the generalizability of our results, we performed the evaluation in two stages:
First, we applied \toolS to all $64$ APIs obtained from APIs.guru (\emph{GuruIndividual} and \emph{GuruGoogleSample}). These APIs do not contain the $15$ APIs we used to train \toolS (see Section~\ref{sec:approach_baseurl}). We decided to use all APIs from APIs.guru in this evaluation because the required manual examination of them is also required to conduct the second research question, and there are limited APIs in APIs.guru to study. 
Second, we performed the evaluation on \emph{HarmonySample}, which is a completely separate data source from APIs.guru. The results for these APIs thus better quantify how well \toolS can potentially generalize to other API documentation pages.
To create \emph{HarmonySample}, we randomly selected $56$ APIs from $560$ APIs obtained from API Harmony for which \toolS was able to generate specifications -- for the remaining $121$ ($681$ -- $560$) APIs, \toolS failed to generate specifications because either no base URLs are described in documentation or \toolS determines all URLs in documentation are not web API invocations by classification (Section~\ref{sec:approach_baseurl}). We limited the number of APIs in \emph{HarmonySample} because the evaluation requires significant manual effort.

Overall, for RQ1, we considered $120$ APIs (\emph{GuruIndividual} + \emph{GuruGoogleSample} + \emph{HarmonySample}). For all of them, we manually identified base URLs, path templates and HTTP methods from web API documentation. We then compared the manually extracted information with the specifications created by \toolS for the same API.
For base URLs, we calculated \emph{precision}, which is the percentage the base URLs generated by \toolS that are correct. Since each API documentation describes only one base URL, and by design \toolS generates one base URL for each API documentation, \emph{recall} is equal to \emph{precision} for base URLs.
For path templates and HTTP methods, we consider \emph{precision} to be the percentage the results generated by \toolS that are correct and \emph{recall} to be the percentage of the given information type (e.g., path templates) in the documentation that \toolS correctly generates. Because path templates and HTTP methods can only be extracted if a base URL was previously detected (see Sections~\ref{sec:approach_paths} and~\ref{sec:approach_methods}), we focus on APIs for which \toolS was able to do so in these parts of the evaluation.

\compactInlineHeader{Results:} \toolS recovered base URLs with a precision of \basePrecisionTotal, inferred path templates with a precision of \pathPrecisionTotal{} and a recall of \pathRecallTotal, and extracted HTTP methods with a precision of \methodPrecisionTotal{} and a recall of \methodRecallTotal. 
Table~\ref{tab:results_rq1} provides a break-down of the results.
\begin{table}[]
\centering
\caption{Precision and recall of \toolS}
\label{tab:results_rq1}
\scalebox{0.72}{
  \begin{tabular}{lcccc}
  \toprule
    & \breakc{\emph{HarmonySample}\\(56 APIs)} & \breakc{\emph{GuruGoogleSample}\\(20 APIs)} & \breakc{\emph{GuruIndividual}\\(44 APIs)} &\breakc{All APIs\\(120 APIs)}\\
\multicolumn{5}{l}{\breakc{ \\ \textbf{Base URL}}}                                         \\ \hline
  Precision    &97.8\%                   & 80.0\%    & 84.1\%          &       87.5\%      \\ \hline\bottomrule

    & \breakc{with correct\\base URLs\\(45 APIs)} & \breakc{with correct\\base URLs\\(16 APIs)} & \breakc{with correct\\base URLs\\(37 APIs)} &\breakc{with correct\\base URLs\\(98 APIs)}\\
\multicolumn{5}{l}{\breakc{ \\ \textbf{Path Template}}}                                         \\ \hline
  \small{\# created D2Spec} & 967 & 188 & 1,219 & 2,374\\ \hline 
  \small{\# in documentation} & 747 & 196 & 1,450 & 2,393\\ \hline 
  \# matches &683 & 187 & 1,060 & 1,930 \\ \hline
  Precision & 70.6\% & 99.5\% & 87.0\%& 81.3\%\\ \hline
  Recall & 91.4\% & 95.4\% & 73.1\% & 80.6\%\\ \hline
\multicolumn{5}{l}{\breakc{ \\ \textbf{HTTP Method}}}                                         \\ \hline
  \small{\# created D2Spec} & 817 & 188 & 1,012& 2,017 \\ \hline 
  \small{\# in documentation} & 815 & 219 & 1,200 & 2,234\\ \hline 
  \# matches &658 & 184 & 861 & 1,703\\ \hline
  Precision & 80.5\% & 97.9\% & 85.1\%& 84.4\%\\ \hline
  Recall & 80.7\% & 84.0\%& 71.8\% & 76.2\%\\ \hline

\bottomrule
\end{tabular}
} 
\end{table}

\subsubsection{Base URL Results}
When manually examining the documentation pages from the $120$ APIs, we found that for eight APIs, the pages did not contain any base URLs. Obviously, \toolS could not generate correct base URLs from such documentation pages and we declared that \toolS is not applicable for such cases. Thus, the eight APIs are excluded in the precision calculation for base URLs extraction. For the remaining $112$ web APIs, \toolS generated correct base URLs for $98$ of them, yielding a precision of \basePrecisionTotal{}.  In the subsequent evaluation for path templates and HTTP methods for RQ1, the evaluation was based on the $98$ APIs.

Upon manual inspection, we found that there were two reasons that \toolS generated incorrect base URLs.
First, when the documentation described multiple API versions, \toolS was unable to tell which one was preferred by the writer of the documentation. For example, in the documentation of the \emph{CityContext web API}, two endpoints were described with \texttt{\footnotesize https://api.citycontext.com/v1/postcodes} and \texttt{\footnotesize https://api.citycontext.com/v2/<location>}. \toolS determined the base URL to be \texttt{\footnotesize https://api.citycontext.com} by selecting the longest common prefix of these two URLs. However, the official documentation listed \texttt{\footnotesize https:// api.citycontext.com/v2} as base URL.
Second, although the classification achieved a good precision, it is unable to remove all URLs that are not web API requests. Such URLs with the same prefix caused \toolS to generate incorrect base URLs when they outnumbered the true web API requests.

\subsubsection{Path Template Results} 
\begin{figure}[t!]
    \centering
    \begin{subfigure}[t]{0.45\textwidth}
        \centering
        \includegraphics[height=1.5in]{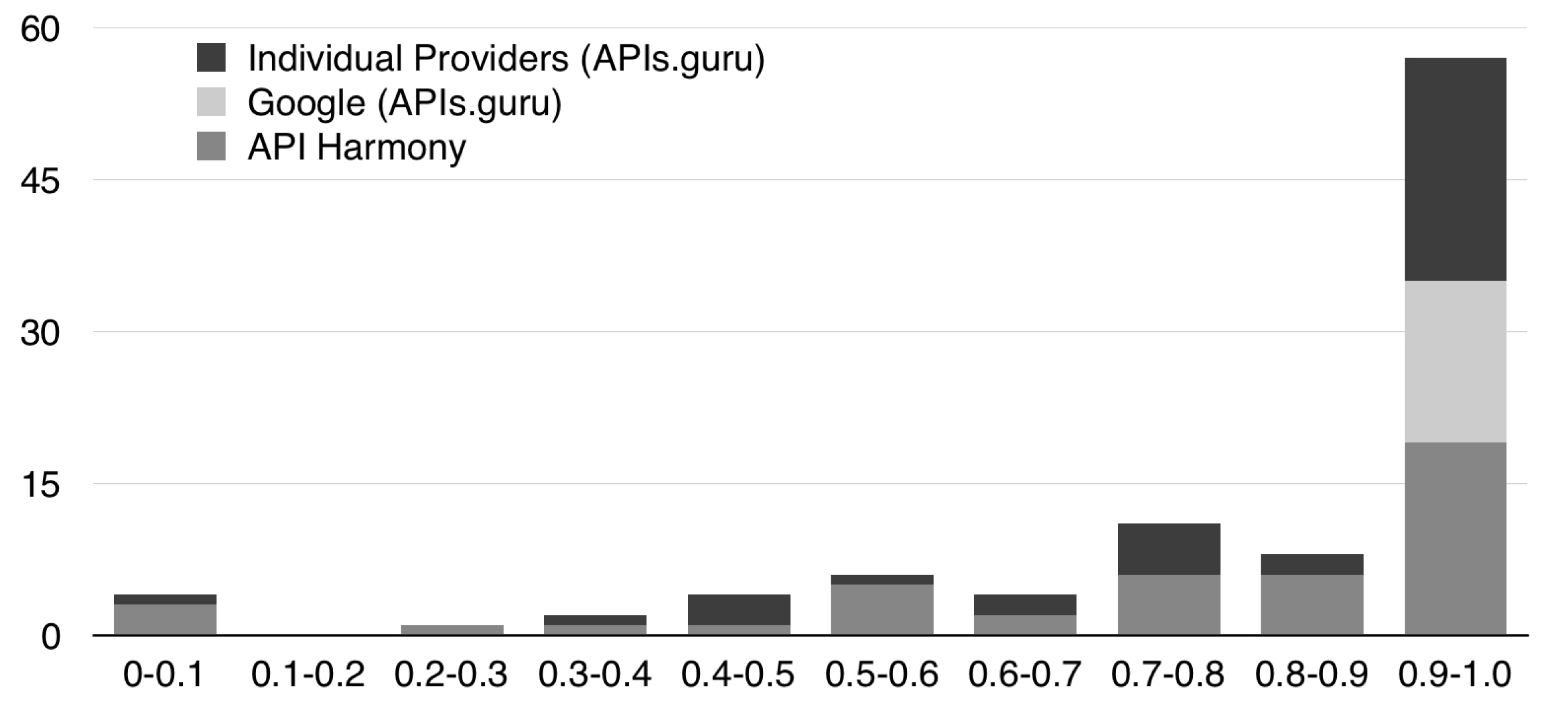}
        \caption{\label{fig:stack_precision}Precision of \toolS on the $98$ web APIs.}
    \end{subfigure}
    \begin{subfigure}[t]{0.45\textwidth}
        \centering
        \includegraphics[height=1.6in]{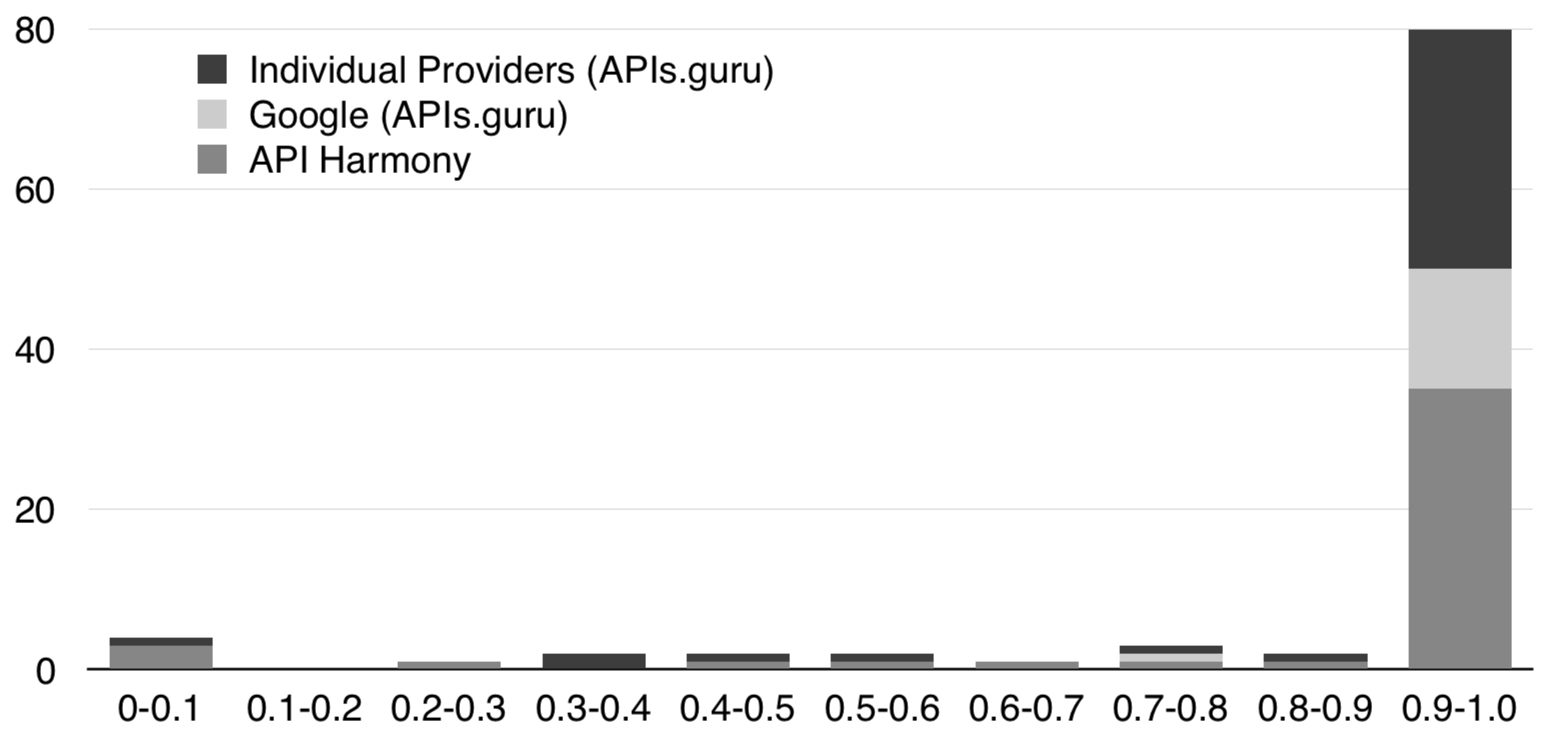}
        \caption{\label{fig:stack_recall}Recall of \toolS on the $98$ web APIs.}
    \end{subfigure}
\caption{\label{fig:stack_overall} Stacked histograms showing precision and recall of \toolS on the $98$ web APIs for which the base URL was correctly extracted.}
\end{figure}

\toolS was able to generate the majority (\pathRecallTotal{} recall) of path templates correctly (\pathPrecisionTotal{} precision) for the $98$ web APIs whose base URLs are correctly identified by \tool. There were in total \pathDocTotal{} path templates described in the documentation. \toolS generated \pathToolTotal{} path templates in total, and \pathToolTotalCorrect{} of them were correct. Thus, the overall precision of path template extraction was \pathPrecisionTotal, and the recall was \pathRecallTotal. Figure~\ref{fig:stack_overall} illustrates stacked histograms on precision and recall of the path template extraction on the $98$ APIs that \toolS can generate correct base URLs for. 
For example, Figure~\ref{fig:stack_precision} shows that for 57 (out of 98) web APIs, \toolS achieves a precision above 90\%.

\subsubsection{HTTP Method Results}
\toolS achieved a precision of \methodPrecisionTotal{} and a recall of \methodRecallTotal{} in extracting HTTP methods for the path templates in the evaluated $98$ web APIs. In total, there were \methodDocTotal{} endpoints with the associated HTTP methods described in the web API documentation evaluated; \toolS produced a result for \methodToolTotal{} of them and \methodToolTotalCorrect{} HTTP methods were correct.
\toolS failed to locate the correct HTTP method when its position in the documentation was far away from the path templates. For example, the \emph{Mandrillapp} documentation has a consolidated description for all endpoints: ``All API calls should be made with HTTP POST'', instead of listing the method \texttt{POST} individually for each of the path template. Thus, \toolS failed to identify correct method names for Mandrillapp's path templates.

\subsection{RQ2: Can \toolS be used to identify inconsistencies between a pre-existing API specification and the API's documentation, pointing to the two being out of synchronization?}

\compactInlineHeader{Approach:} We focused on the $64$ APIs from APIs.guru (\emph{GuruIndividual} + \emph{GuruGoogleSample}). For these APIs, we compared the specifications generated by \toolS (from hereon denoted as \textbf{\emph{ToolSpecs}}) with the specifications provided by APIs.guru (from hereon denoted as \textbf{\emph{GuruSpecs}}). Our comparison focused, again, on the three pieces of information extracted by \tool, namely, base URLs, path templates, and HTTP methods.
For \textbf{base URLs}, we compared whether the ones extracted by \toolS per API match those defined in the OpenAPI specifications. We obtained base URLs from the specifications by concatenating the \texttt{schemes}, \texttt{host}, and \texttt{basePath} fields.
We then compared whether the extracted \textbf{path templates} and the associated \textbf{HTTP methods} match the ones in the specifications.

For each of the three extracted pieces of information, we counted the number of matches, and then manually inspected the mismatches to determine their origin.

\compactInlineHeader{Results:} We found that mismatches between \emph{GuruSpecs} and \emph{ToolSpecs} were partly caused by limitations of \tool, and partly by publicly-available specifications (i.e., the \emph{GuruSpecs}) being out of synchronization with API documentation:
Our manual inspection showed that for base URLs and HTTP methods, \emph{GuruSpecs} were often up-to-date with documentation, and mismatches between \emph{ToolSpecs} and \emph{GuruSpecs} were due to inaccuracies of \tool.
However, for path templates, our manual inspection found that many mismatches were due to the documentation and \emph{GuruSpecs} being out of synchronization with each other, or due to errors in the documentation. Specifically, for the $64$ APIs evaluated, we identified $385$ path templates from $21$ APIs where \emph{GuruSpecs} and the documentation were different. One reason for the mismatches is that as web APIs evolve, API providers tend to keep documentation up-to-date since it is, as a human-readable medium, often the first source that developers inquire to use APIs.
In the following, we present the results of manually examining the mismatches between path templates in \emph{GuruSpecs} and \emph{ToolSpecs}. We found that mismatches fall into four categories:

\begin{itemize}
  \item \textbf{Inconsistencies} were mismatches resulting from the documentation and specification in \emph{GuruSpecs} being inconsistent with each other. Such inconsistencies were not caused by deficiencies of \toolS or by errors in the documentation, but indicated that the API provider should either update the documentation or the specification. For example, in the documentation of \emph{Slack}, there were eight endpoints on getting information on members from a given Slack team. The paths of all eight endpoints start with {\tt /users.<action>}. However, only one path template was listed in the specification--{\tt /users.list}; the remaining seven (e.g., {\tt /users.info} and {\tt /users.setPresence}) were missing in the specification. In this case, we considered that there were seven inconsistencies between the documentation and the specification.

  \item \textbf{Errors in the documentation} referred to obvious errors in the documentation, e.g., typos. Incorrect information in the documentation led \toolS to generate path templates that, while being labeled as correct with regard to RQ1, did not match the specifications. For example, in the documentation of the \emph{ClickMeter API}, many path templates starting with {\tt /data{\bf t}points} were misspelled as {\tt /datapoints}. Thus, \toolS generated several mismatched path templates compared to the official specification.

  \item \textbf{Partially correct path templates} occurred if \toolS failed to infer path parameters correctly (i.e., the path templates generated by \toolS still contain path parameter values). A common reason for this problem was that the documentation contained only one instance of an instantiated value for a path parameter. In such cases, even though \tool's clustering-based algorithm can correctly place the path in its own cluster, \toolS could not distinguish which segments of the path were instantiated values and which ones are fixed segments.

  \item \textbf{Deficiencies in the algorithm} occurred when \toolS failed to extract certain path templates or generated incorrect path templates because of deficiencies in its design. For instance, \toolS failed to extract certain path templates if the way the path templates appeared in the documentation was beyond the scope of the conventions used by \tool. \toolS relies on the format of URL and relative path to extract path template information. If the path templates are not presented as such, \toolS will not extract them correctly. For example, \emph{HealthCare.gov}'s documentation describes a series of path templates as follows: ``The following content types are available: articles, blog, questions, glossary, states, and topics. The request structure is https://.../api/:content-type.json.'' \toolS extracts one path template--{\tt /api/\{content-type. json\}} instead of six path templates (e.g., {\tt /api/articles} and {\tt /api/blog}). On the other hand, using the conventions mentioned above, \toolS also extracted false path templates which did not describe path templates in the documentation. For example, the documentation of \emph{dweet.io} listed a file path -- {\tt /play/definitions}, which was not a true path template.
\end{itemize}

\begin{figure}
\scalebox{0.25}{
\includegraphics[]{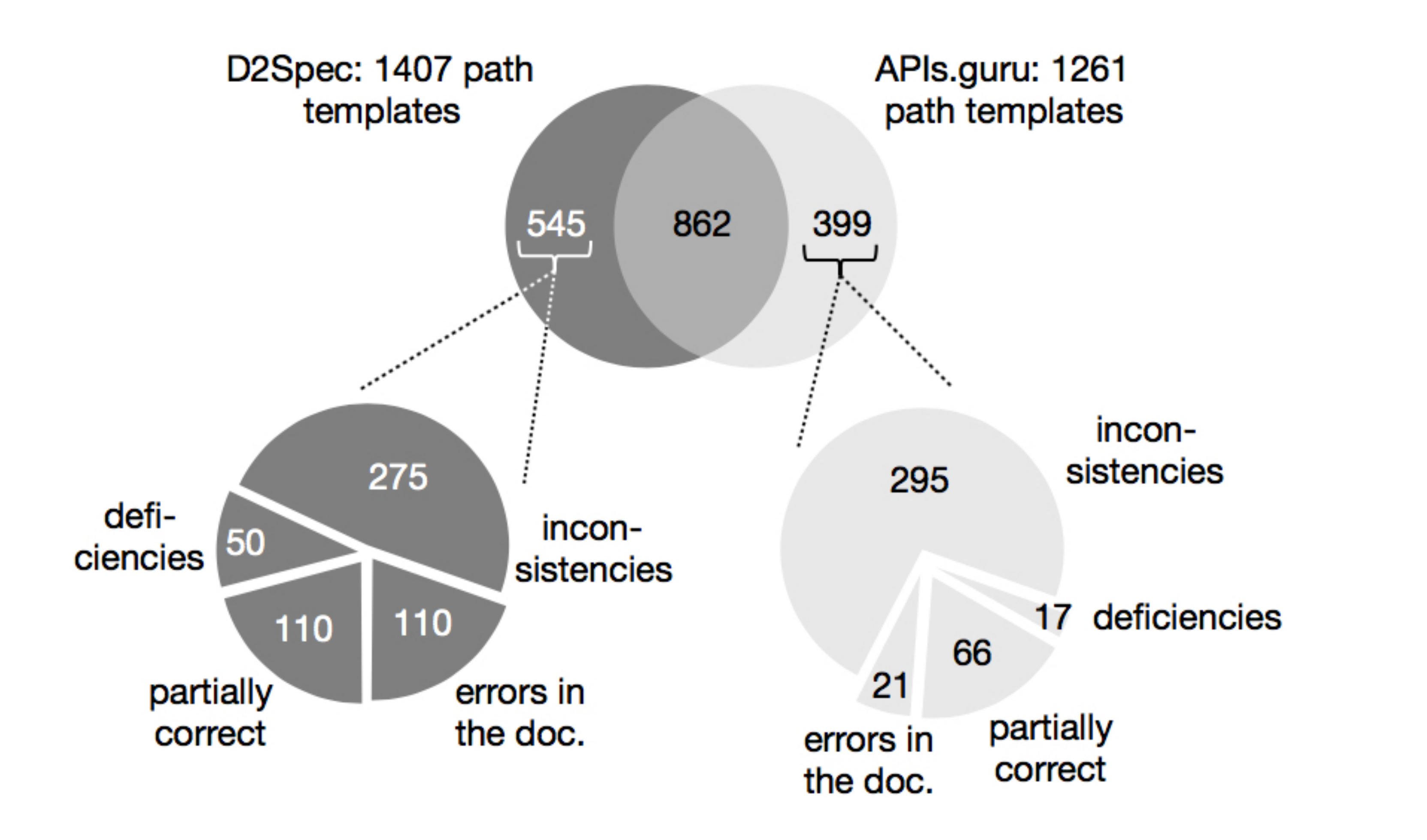}
}
\caption{Comparison between path templates in specifications from \toolS and the ones from APIs.guru.\label{fig:comparison}}
\end{figure}

Figure~\ref{fig:comparison} visualizes the comparison of path templates from \emph{GuruSpecs} and \emph{ToolSpecs}. The breakdown of the mismatches from both aspects ($\neg\emph{ToolSpecs} \cap \emph{GuruSpecs}$ and $\emph{ToolSpecs}\cap\neg\emph{GuruSpecs}$) is shown as well. In total, there are $862$ path template matches between \emph{ToolSpecs} and \emph{GuruSpecs}. Among the $1407$ path templates generated by \tool, there are $545$ mismatches with the path templates defined in the \emph{GuruSpecs}. Our manual analysis shows that $385$ ($70.6\%$) of the mismatches are caused by de-synchronization, i.e., ``inconsistencies'' and ``errors in the doc.''. The other two categories -- ``partially correct'' and ``deficiencies'' -- are due to limitations of \tool.

\begin{figure}
\includegraphics[height=2.3in]{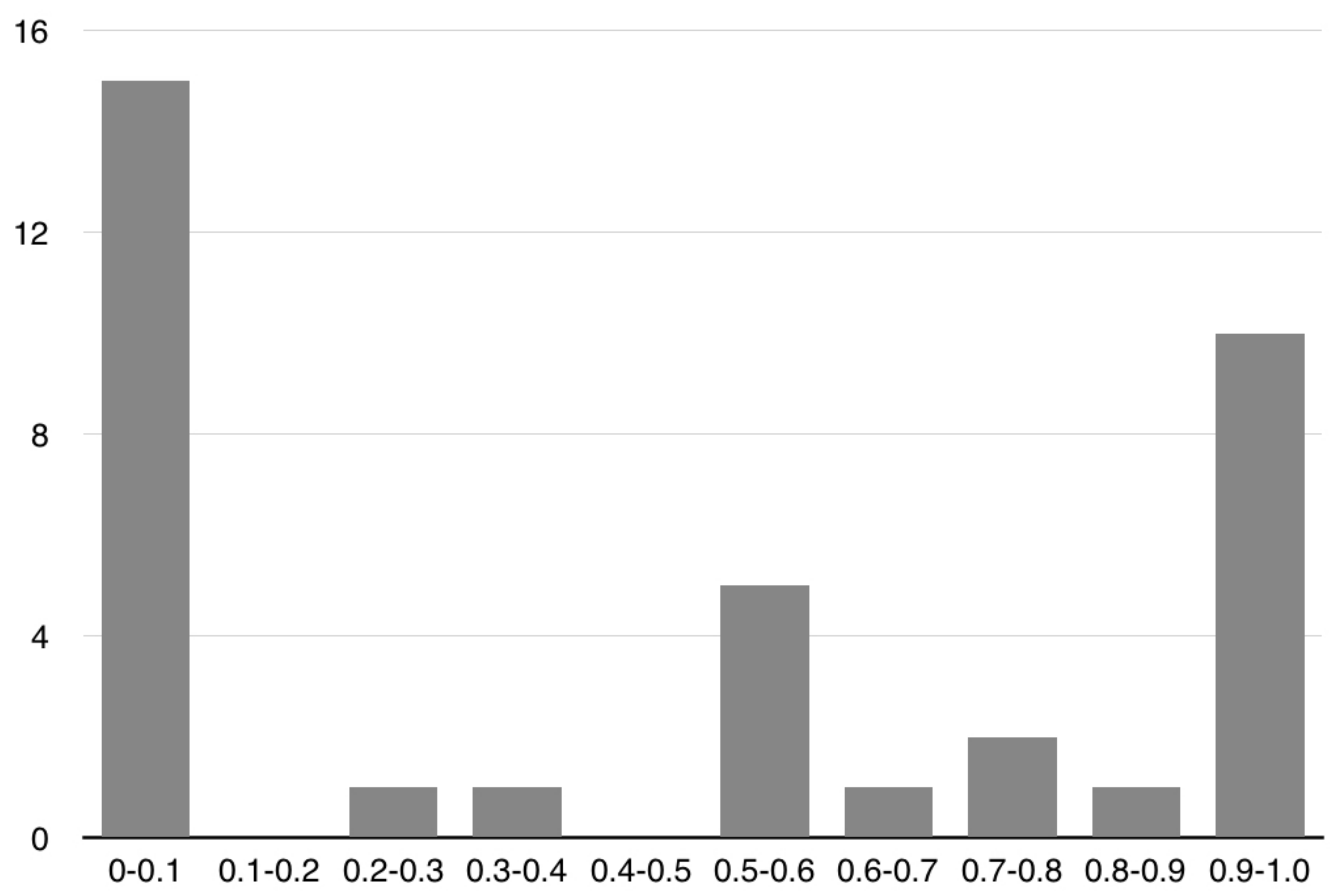}
\caption{Percentage of ``Inconsistencies'' and ``Errors on the documentation'' across APIs from APIs.guru\label{fig:stack_mismatch}}
\end{figure}

Overall, while the manual examination of mismatches also pointed to some weaknesses of \tool, it also highlights that \toolS can be used to find documentation and existing specifications being out of synchronization. To focus on this aspect, Figure~\ref{fig:stack_mismatch} shows a histogram on the percentage of mismatches that are caused by de-synchronization for each web API. It shows that, for example, for $10$ web APIs, over $90\%$ of the mismatches detected by \toolS indicate that documentation and pre-existing specifications were out of synchronization with each other.

\medskip

\section{Threats to Validity and Discussions}
\label{sec:threats}

\noindent\textbf{Generalizability}. \toolS generates base URL from documentation by firstly identifying URLs that represent web API calls through a classification algorithm. The classification algorithm uses a set of pre-labeled URLs for training. We built the training set from a set of web API documentation that were independent of the ones used in the evaluation. The precision of \toolS in base URL extraction may be different if we use a different training set. However, we mitigated this bias by choosing a random set of web APIs for building the training set, and evaluating on a different set of APIs (the \emph{GuruIndividual} and \emph{GuruGoogleSample} datasets described in Section~\ref{sub:data}). In addition, we evaluated on a completely separate dataset (\emph{HarmonySample}) and even achieved a better precision (97.5\%) compared to a 80.0\% precision on \emph{GuruGoogle} and a 84.1\% precision on \emph{GuruIndividual}, demonstrating that our approach is likely to generalize to other unseen documentation pages. 

\noindent\textbf{Thresholds Used in the Clustering-Based Path Template Extraction}. \toolS leverages an iterative clustering-based algorithm to identify path parameters by inferring values of path parameters from \textit{similar} web API calls. The proposed algorithm contains thresholds to control the hierarchical clustering (e.g., determining whether two web API calls are \textit{similar} through a threshold \textit{T}, see Algorithm~\ref{clustering} in Section~\ref{sec:approach_paths}). In this evaluation, we set the thresholds based on our observations on the training set. We found that the chosen thresholds also worked well for the evaluation set. Nevertheless, future studies should investigate the effects of different thresholds on the path template extraction results. 

\noindent\textbf{Documentation with Identical Structures}. APIs from the same providers may have identical documentation structures (e.g., Google web APIs). Documentation structures may be different across different API providers. 
To show the generalizability of our approach, we applied our approach to APIs from different providers: Our evaluation set contains 120 APIs from 98 different web API providers. 

\section{Related Work}
\label{sec:related_work}

We discuss related work in using information extraction approaches in obtaining software or other types of entities, and other methods of inferring web API specifications.

Many software engineering researchers have looked into the problem of identifying code elements---more specifically, Java code elements such as method signatures and calls---from API documentation.
Dagenais and Robillard proposed an approach that extracts code elements from API documentation and links the elements to an index of known code elements, i.e., signatures from a Java library~\cite{recover2012}.  
Subramanian et al. subsequently applied this approach to identify code elements on Stack Overflow posts and augmented the code elements in the posts with links to their official JavaDoc~\cite{subramanian2014live}. 
Rigby and Robillard use a light-weight, regular expression based approach to identify code elements that relaxes the requirement on a known index~\cite{rigby2013discovering}.
Another line of work focuses on extracting more complex specifications on code entities from natural language descriptions. Pandita et al.~\cite{pandita2012} extract method pre-conditions and post-conditions from natural language API documentation.
Lin et al.~\cite{icomment} extract code contracts from comments and statically check for violations in the code.
Our work differs in two ways. First, we extract web API endpoints and related information as opposed to code elements. Second, there is arguably greater value in our recovered index (i.e., OpenAPI Specifications) because such an index is often not available or known to the clients; while clients of Java libraries (or other statically-typed languages) are always exposed at least to method signatures, but callers of web APIs often do not have such information.

There have been many efforts in information extraction on web pages~\cite{alvarez2008extracting,crescenzi2001roadrunner,ferrara2014web,laender2002brief,myllymaki2002effective,zhai2005web,zhai2007extracting}. 
For example, techniques for extracting product information from e-commerce sites~\cite{zhai2005web,zhai2007extracting} leverage the structure from the sites: the sites' organizational structure usually consists of a search page and a set of individual product pages, which typically have the same structure as they are generated from scripts.  These techniques exploit this common structure across the pages within the same site.
However, for extracting endpoints and other information from web API documentation pages, we cannot rely on such an assumption: There is no standard structure for API documentation.
For many API documentation the content is semi-structured at best, written by humans using free-form text and/or diverse HTML structures.
For example, the GitHub API documentation uses an example-based style, where the base URL \texttt{https://api.github.com} and the path template \texttt{/users/\{username\}/orgs} are embedded in free-form text and a \texttt{curl} command.
Other documentation uses a more structured, reference-based documentation style.

There have been other attempts at extracting OpenAPI Specifications. Wittern and Suter use dynamic traces in form of web-server logs~\cite{suter2015inferring}. Or, the SpyREST tool intercepts HTTP requests to an API using a proxy and then attempts to infer the API documentation~\cite{Sohan:2015dp}. Our work differs in that we use a different source of input, namely API documentation, as opposed to dynamic traces or observed traffic. The benefit of our approach is that API documentation is meant to be public while access to web logs are limited to those with access to the private web servers, and proxying may not be an option.

\section{Conclusion}
\label{sec:conclusion}
In this paper, we presented \tool, a tool which extracts parts of web API specifications from documentation, including base URLs, path templates, and HTTP methods. \toolS is based on the three assumptions: (1) documentation includes multiple web API URLs (so that a base URL can be extracted); (2) path templates are either denoted explicitly (e.g., using brackets) or that multiple example URLs for paths exist from which templates can be inferred; and (3) descriptions close to the path templates contain information about HTTP methods. Our evaluation of \toolS shows that these assumptions hold mostly true when it comes to extracting base URLs, path templates, and HTTP methods. It furthermore shows that \toolS is not only useful for creating specifications from scratch, but also for checking existing ones for consistency with documentation. In the future, we aim to expand the scope of \toolS to also extract information on data structures, HTTP headers, and authentication methods.
\balance

\bibliographystyle{abbrv}



%

\end{document}